\documentclass[12pt]{article}

\usepackage[normalem]{ulem}
\usepackage[mathscr]{eucal}
\usepackage[version=3]{mhchem}

\usepackage{epsfig,latexsym,amsfonts,amsmath,amsthm,amssymb,amsbsy,multirow,slashed,wasysym,textcomp,dsfont,comment,mathtools,cancel,cite,datetime,appendix,BOONDOX-calo}
\usepackage{tocloft}
\usepackage{bbold}
\usepackage{tikz}
\usetikzlibrary{backgrounds}
\usetikzlibrary{patterns}
\usetikzlibrary{arrows.meta}
\usetikzlibrary{shapes,snakes}
\tikzstyle{bag} = [align=center]
\usetikzlibrary{decorations.pathmorphing}
\def\bea{\begin{eqnarray}}
\def\eea{\end{eqnarray}}

\usepackage{hyperref}
\usepackage{subcaption}

 \newcommand{\badat}{\begin{alignedat}}
 \newcommand{\eadat}{\end{alignedat}}
 
 \def\be{\begin{equation}}
\def\ee{\end{equation}}

\def\p{\partial}

\usepackage{xcolor}

\newcommand{\pink}[1]{\textcolor{\pink}{#1}}

\definecolor{dblue}{rgb}{0.2,0.50,0.80}

\def\bz{{\bar z}}

\def\bz{{\bar z}}

\def\scri{\mathcal I}

\DeclareFontFamily{OT1}{pzc}{}
\DeclareFontShape{OT1}{pzc}{m}{it}{<-> s * [1.10] pzcmi7t}{}
\DeclareMathAlphabet{\mathpzc}{OT1}{pzc}{m}{it}

\definecolor{vert}{rgb}{0.1367 0.543 0.1367}

\numberwithin{equation}{section} 

\begin{document}

 \begin{titlepage}
  \thispagestyle{empty}
  \begin{flushright}
  \end{flushright}
  \bigskip
  \begin{center}

        \baselineskip=13pt {\Large \scshape{
      A Shorter Path to Celestial Currents  
      }}
       
      \vskip1cm 

   \centerline{ 
   {Sabrina Pasterski}
}

\bigskip\bigskip
 
 \centerline{
 Princeton Center for Theoretical Science, Princeton, NJ 08544, USA}

\bigskip\bigskip

\end{center}

\begin{abstract}
 \noindent 
 
Here we consider what happens when we lift a codimension-1 slice of the celestial sphere to a codimension-1 slice of the bulk spacetime  in a manner that respects our ability to quotient by the null generators of $\mathcal{I}^\pm$ to get to our codimension-2 hologram. The contour integrals of the 2D currents for the celestial symmetries lift to the standard boundary integrals of the 2-form generators for the gauge theory and celestial Ward identities follow directly from Noether's theorem.

\end{abstract}

\end{titlepage}

\tableofcontents



\section{Introduction}
Asymptotically flat spacetimes present a surprisingly rich symmetry structure, with infinite dimensional symmetry enhancements appearing when we consider gauge transformations that act non-trivially on the asymptotic data.  As elucidated by Strominger~\cite{Strominger:2017zoo} these give rise to non-trivial Ward identities manifested as soft theorems for $\mathcal{S}$-matrix elements.

However in repeating the standard derivation of the `Ward identity $=$ soft theorem' equivalence, one can't help but take pause at the number of steps needed to demonstrate such an elegant final result. Nominally we would proceed roughly as follows
\begin{enumerate}
    \item Take a Cauchy slice of my asymptotically flat spacetime.
    \item Push it down to past null infinity to define my $in$-state.
    \item Push it up to future null infinity to define my $out$-state.
    \item Evaluate the canonical charges on each slice.
    \item Use an antipodal matching condition (across infinitely time-separated spheres!)  at $\mathcal{I}^+_-$ and $\mathcal{I}^-_+$ to equate these two expressions.
    \item Use the constraint equations and integrate by parts to write this charge as a flux.
    \item Insert this as an operator equation in $\mathcal{S}$-matrix elements and see that 
    \be
    \langle out|(Q^+_{S}+Q^+_H)\mathcal{S}-\mathcal{S}(Q^-_{S}+Q^-_H)|in\rangle=0
    \ee
    arises from the soft theorem by plugging the perturbative mode expansion into $Q_S$, which is linear in the gauge field.
\end{enumerate}
This summary does not include the additional historical step where we would ignore massive contributions in the early derivations~\cite{He:2014laa,Kapec:2014opa,He:2014cra}, to make our lives easier when integrating by parts along $u$ and $v$ in step 6. 

These Ward identities play a crucial role in motivating the celestial holographic dictionary, whereby $\mathcal{S}$-matrix elements in a boost basis transform like correlators in a CFT living on the celestial sphere~\cite{Pasterski:2016qvg,Pasterski:2017kqt,Pasterski:2017ylz}. It thus strongly behooves us to have a clean bulk interpretation of the manipulations we are doing in 2D. The aim of this note is to show that we can skip to what is roughly step 6 at least for the purposes of celestial Ward identities if we combine the  Noether's second theorem-based approach of~\cite{Avery:2015rga}, with an extension of the left/right Hilbert space picture of~\cite{Crawley:2021ivb} into the bulk, and the extrapolate dictionary of~\cite{Pasterski:2021dqe}.

The punchline of our story is that for an appropriate choice of 3-surface $\Sigma_C$ the celestial charge $\mathcal{J}_C$ in the radially quantized CCFT is related to the surface charge of the corresponding gauge theory in the bulk
\be\label{result}
\boxed{~~\mathcal{J}_C(\lambda)=\int_{\p\Sigma_C}\star k(\lambda).~~}
\ee
Namely we can jump directly from the canonical charges in the bulk to the symmetry generators in the boundary in a manner similar to what one would do in AdS/CFT. The price we pay is that the surface $\Sigma_C$ is not a Cauchy slice, but this is not surprising considering that radial evolution is Rindler time evolution which is spacelike outside the  Rindler wedge. In terms of the standard presentation, the bulk operator corresponding to the left hand side is normally constructed from the soft charge. By deriving~\eqref{result} we get an immediate route to it. As we will explore in more detail below this should help explain why the soft and hard splitting of the flux operators separately obey the symmetry algebra relations.

\section{Surface Charges for Gauge Symmetries}

In this section we will follow the conventions of~\cite{Avery:2015rga}.   We will focus on the electromagnetic case for simplicity and use units where $e=1$. 

Recall that in gauge theory we can write the canonical charges for a gauge transformation $\lambda$ in terms of a 2-form $k(\lambda)$. For a Cauchy slice $\Sigma$ we have
\be
Q(\lambda)=\int_{\p\Sigma} \star k .
\ee
For the case of electromagnetism the charge generating the gauge transformation
\be
A_\mu\mapsto A_\mu+\p_\mu\lambda 
\ee
corresponds to 
\be\label{klF}
k=\lambda F,~~~F=\frac{1}{2}F_{\mu\nu}dx^\mu\wedge dx^\mu
\ee
so that on a constant time Cauchy slice this reduces to 
\be
Q(\lambda)=\lim_{r\rightarrow\infty} \int_{S^2} d^2z\sqrt{\gamma}~\lambda(r^2 F_{ru})
\ee
where we use the coordinates
\be
X^\mu=(u+r,r\frac{z+\bz}{1+z\bz},ir\frac{\bz-z}{1+z\bz},r\frac{1-z\bz}{1+z\bz}).
\ee
Large gauge transformations are precisely those for which $\lambda$ has non-compact support so that it survives this large $r$ limit.  In the case where $\lambda$ is constant we get back to the global $U(1)$ symmetry generator, but in the context of studying asymptotic symmetries we allow $\lambda(z,\bz)$.

On the equations of motion we have
\be\label{ktoj}
\p_\nu k^{\nu\mu}\overset{w}{=}{\cal j}^\mu~~~~ \Rightarrow ~~~~\int_{\p R} \star k=\int_R \star {\cal j}
\ee
where ${\cal j}$ is the Noether current one gets from path integral Ward identities.  For our electromagnetic example
\be\label{jem}
{\cal j}_\mu=F_{\nu\mu}\p^\nu\lambda+\lambda J^M_\mu
\ee
where the last term is the matter current.  The right hand equality in~\eqref{ktoj} is the starting point for talking about charge conservation.  Namely, if we consider two Cauchy slices with the same boundary, the charges would be the same.

The essence of the asymptotic symmetry Ward identities is to write 
\be
\p M=\Sigma_1\cup\Sigma_2
\ee
and try to take $\Sigma_1\mapsto \scri^-$ and  $\Sigma_2\mapsto \scri^+$. The series of steps needed in our outline is to justify that there is no additional contribution near spatial infinity $i^0$. Namely
\be
\p \mathcal{I}^{\pm}= \mathcal{I}^{\pm}_+\cup \mathcal{I}^{\pm}_- 
\ee
 and until we have the antipodal matching condition it is not obvious that $k|_{\mathcal{I}^+_-}=k|_{\mathcal{I}^-_+}$.  Once we do have this matching condition, the Ward identity has a very nice interpretation in terms of a relationship between time integrals of radiation called memory effects and the external scattering states~\cite{Pasterski:2015zua}.  By~\eqref{jem} the charge at future null infinity is
\be\label{Qp}
Q^+=\lim\limits_{r\rightarrow\infty} r^2\int du\sqrt{\gamma} d^2 z n^\mu(F_{\nu\mu}\p^\nu\lambda+\lambda J^M_\mu)
\ee
where the normal vector of null infinity is
\be
n^\mu\p_\mu =\p_u-\frac{1}{2}\p_r.
\ee
For the standard radiative phase space only the $F_{uA}$ term contributes. Moreover, each of these terms is $\mathcal{O}(1)$ in the large-$r$ limit since raising the sphere metric gives us an $r^{-2}$ in the first term. The first term in~\eqref{Qp} is the soft charge $Q^+_S$ while the second term is the massless contribution to the hard charge $Q^+_H$. Step 6 in our procedure is precisely the fact that the weak equality in~\eqref{ktoj} amounts to using the constraint equations. For the full Ward identity we need to repeat this procedure for past null infinity and take into account additional contributions near $i^{\pm}$ for the massive scatterers~\cite{Campiglia:2015qka,Kapec:2015ena}.  A nice recent review can be found in~\cite{Miller:2021hty}.  This procedure generalizes to any other asymptotic gauge symmetry.  The appearance of a term in the charge that is linear in the gauge field is symptomatic of the fact that the majority of these asymptotic symmetries are spontaneously broken by our choice of vacuum. The Ward identities are manifested in $\mathcal{S}$-matrix elements as soft theorems precisely because the term linear in the gauge field reduces to a $u$-integral that picks out a zero frequency mode of the gauge field~\cite{He:2014cra}.

\section{Celestial Symmetries}

The main motivation for the celestial CFT program is that these Ward identities for 4D asymptotic symmetries can be naturally recast as 2D ward identities for a CFT living on the celestial sphere. This is most striking for the case of the subleading soft graviton~\cite{Cachazo:2014fwa,Kapec:2014opa}, from which we get a candidate 2D stress tensor~\cite{Kapec:2016jld} if we go to a basis of boost-eigenstates.  For the U(1) gauge theory case at hand we can define the following current~\cite{Strominger:2013lka,He:2015zea,Nande:2017dba}
\be\label{jisqs}
j=j^++j^-,~~~~j^+=Q_S^+\left(\lambda=\frac{1}{z-w}\right)
\ee
where $j^-$ is a corresponding antipodally-related $\mathcal{I}^-$ contribution.  Because of a helicity redundancy~\cite{He:2014cra} (corresponding to a shadow relation for $\Delta=1$ photons~\cite{ss}) the $F_{uz}$ and $F_{u\bz}$ contributions to~\eqref{Qp} are equal, and we can just as well write
\be\label{jfuz}
j^+=-{4\pi}\int du F_{uz}.
\ee
Viewing $\mathcal{S}$-matrix elements as celestial correlators, Weinberg's soft photon theorem~\cite{Weinberg:1965nx} turns into the relation 
\be\label{jzward}
\langle j(z)\mathcal{O}_1(z_1,\bz_1)...\mathcal{O}_n(z_n,\bz_n)\rangle=\sum_i\frac{Q_i}{z-z_i}\langle\mathcal{O}_1(z_1,\bz_1)...\mathcal{O}_n(z_n,\bz_n)\rangle.
\ee
We can generate an arbitrary holomorphic $U(1)$ transformation $\lambda$ using the operator
\be\label{Jcvj}
\mathcal{J}_C(\lambda)=\oint_C\frac{dz}{2\pi i }\lambda j
\ee
so that 
\be
\langle \mathcal{J}_C(z)\mathcal{O}_1(z_1,\bz_1)...\mathcal{O}_n(z_n,\bz_n)\rangle=\sum_{i\in C}{Q_i}\lambda(z_i)\langle\mathcal{O}_1(z_1,\bz_1)...\mathcal{O}_n(z_n,\bz_n)\rangle.
\ee
In particular we get the global transformations when we take $\lambda=1$ and the contour $C$ to enclose all of the charged scattering states. In that case the left hand side vanishes, which we can see by shrinking the contour in the direction where there are no operator insertions. We see this from point of view of the soft charge by the fact that $Q_S$ vanishes when $\p_\mu\lambda=0$.

We emphasize that we want $\lambda$ holomorphic (not just meromorphic) within the contour $C$. As pointed out in~\cite{He:2015zea} if $\lambda$ has poles within this contour then we will get extra insertions of $j$ coming from the residues. It is in this sense that $\mathcal{J}$ is more properly thought of as the full charge rather than just the soft contribution. Within $\mathcal{S}$-matrix insertions this was appreciated in~\cite{He:2015zea}, however their starting point was at the level of the soft theorems. 
In what follows we will be able to give an extra dimension (pun intended!) to the statements about contours dividing in and out particles we see in~\cite{He:2015zea} and the cover art of~\cite{Strominger:2017zoo}.

\section{Slicing the Sphere and the Bulk}

Here we will show that from the canonical charge perspective it is clear that the quantity $\mathcal{J}_C$ is the object generating gauge transformations on the operators within the contour. The essence of what we will be doing in this section is to consider what happens when we do the alternate splitting 
\be
\p M=\Sigma_1\cup\Sigma_2 = \Sigma_L\cup \Sigma_R
\ee
where the contour $C$ that divides the left from the right patch of the celestial sphere determines the division between left and right patches of our boundary. The downside is that these left and right surfaces are not Cauchy slices.  However, this splitting is actually natural from the point of view of celestial CFT.

The main reason for going to the celestial basis is so that we can better take advantage of the asymptotic symmetries that should constrain scattering. The structure of the celestial OPEs derived from collinear limits of scattering~\cite{Fan:2019emx,Fotopoulos:2019tpe,Pate:2019lpp,Fotopoulos:2019vac} strongly suggest that we can treat them as if we are doing radial quantization.  In doing so one finds an even more surprisingly rich symmetry structure~\cite{Guevara:2021abz,Strominger:2021lvk,Himwich:2021dau}.
Recent attempts to formalize the out states and the state-operator correspondence in this language can be found in~\cite{Fan:2021isc,Crawley:2021ivb}.  Here we will be particularly interested in the left/right splitting of states suggested in~\cite{Crawley:2021ivb}.

\begin{figure}[ht]
    \centering
    \includegraphics[trim=150 100 100 50,clip, width=15cm]{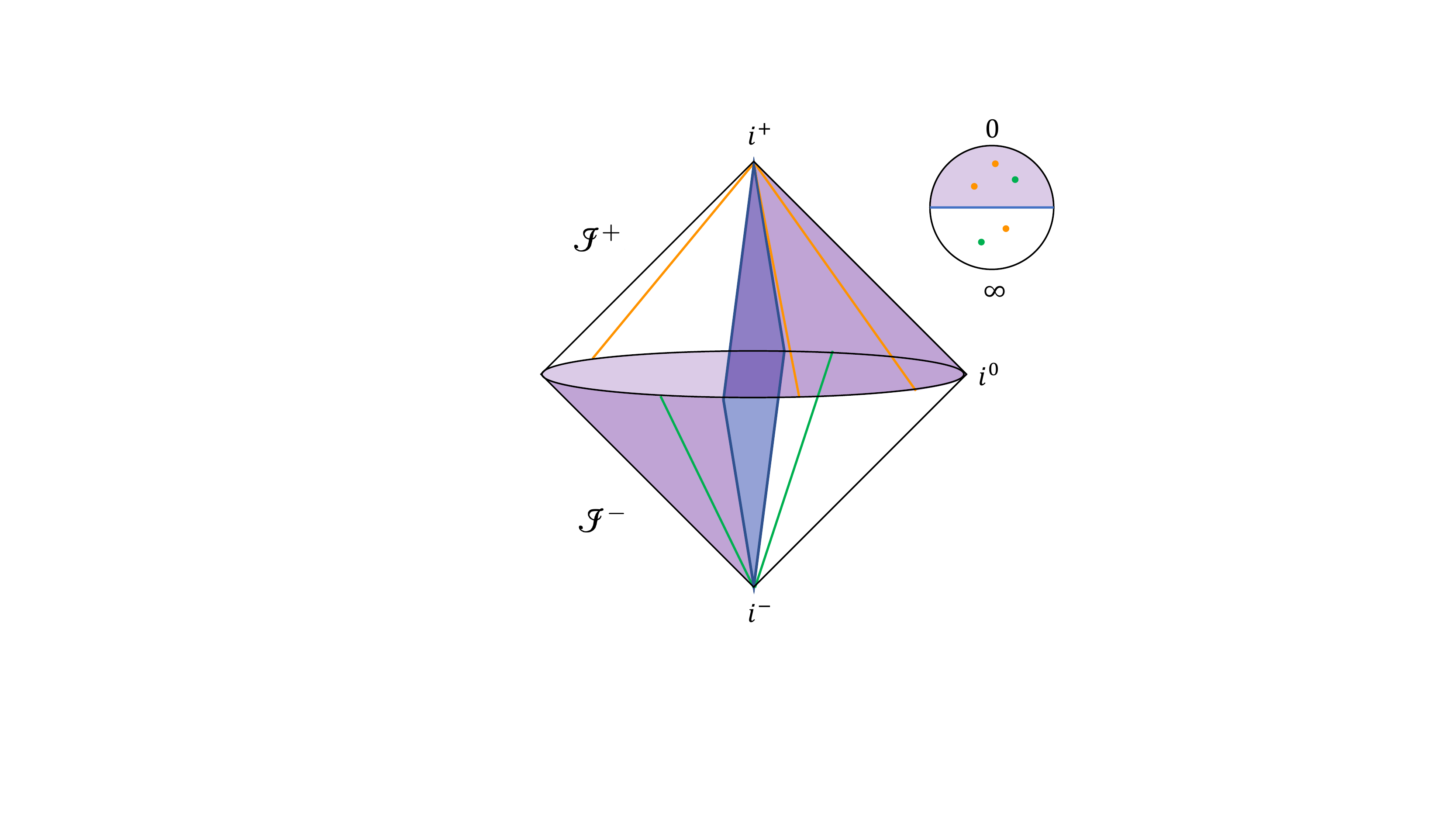}
    \caption{Splitting the spacetime across the hyperplane $X^3=0$ splits the celestial sphere along the equator. By the extrapolate dictionary, we can prepare massless scattering states by inserting operators along generators of null infinity. The north patch of the celestial sphere corresponds to the regions shaded in purple, related across spatial infinity by an antipodal matching.}
    \label{slice}
\end{figure}

Radial evolution in celestial CFT corresponds to Rindler evolution in the bulk.\footnote{In~\cite{Pasterski:2022lsl} we examine of some implications for CCFT that naturally arise when considering the perspective of a Rindler observer. Here the direction of the particle's acceleration sets the foliation of both the sphere and the bulk.}  If we slice our spacetime along the hypersurface $X^3=0$, the celestial sphere will get cut along its equator $|z|=1$.  This is illustrated by the blue surface in figure~\ref{slice}.  In a reference frame boosted along the $X^3$ direction this slice will tilt towards one or the other Rindler horizons and the corresponding locus on the celestial sphere will shrink.  Global conformal transformations of the Riemann sphere map circles to circles.  Global conformal transformations of the celestial sphere are induced by Lorentz transformations in the bulk. Under such Lorentz transformations we can go from our $X^3=0$ hyperplane to any other hyperplane with spacelike normal. As such we expect to be able to map any circle on the celestial sphere to a hyperplane with spacelike normal in the bulk.\footnote{We've seen the special role of `celestial circles' popping up in Mellin-transformed momentum space when examining the 4-pt kinematics for massless scattering~\cite{Pasterski:2017ylz}. We will show that this hyperplane trick has a much more general utility when studying constraints from translation invariance on celestial amplitudes in~\cite{MP}.}  While more intricate contours $C$ will not be boost images of this canonical contour, one can consider propagating into the bulk the deformation one would do on the celestial sphere to get from a circle to $C$.

What is nice about this hypersurface is that it respects our ability to quotient along the generators of null infinity. When we go down by one codimension to a Cauchy slice, we have a boundary that is the full sphere at infinity.  While this is a codimension-1 cut of the boundary if we want to look at $u$-evolution, we have to do a lot of gymnastics to construct an object that looks like a radial quantization `time'-slice. Namely the steps outlined in the introduction. Instead our hypersurface $\Sigma_C$ extends along the $u$-direction and is also codimension-1 on the celestial sphere.

Now the boundary of this slice is still in the asymptotic region and that is all that we will need to evaluate the charges $Q^C$.  Recalling that
\be
\star(dx^\mu \wedge dx^\nu)=\sqrt{-g}g^{\mu\kappa}g^{\nu \lambda}\epsilon_{\kappa\lambda\rho\sigma}\frac{1}{2!}dx^\rho\wedge dx^\sigma
\ee
and using the radiative falloffs
\be
F_{ur}\sim\mathcal{O}(r^{-2}),~~~F_{z\bz}\sim\mathcal{O}(1),~~~F_{uA}\sim\mathcal{O}(1),~~~F_{rA}\sim\mathcal{O}(r^{-2})
\ee
the integral of the 2-form~\eqref{klF} evaluates to 
\be\label{QFC}
Q^C(\lambda)=\int_{\p \Sigma_C}\star (\lambda F)= -i \int du \oint_C dx^A \lambda \epsilon_{AC}\gamma^{CB}F_{ B u} + (\mathcal{I}^-~\rm{contribution})
\ee
(where $\epsilon_{AC}$ contains a factor of $\sqrt{\gamma}$) which we recognize as a contour integral of the memory operator, and which is determined by Weinberg's soft photon theorem.  While in the standard presentation the `weak' equality we need to go from from the charge as a surface integral to the flux presentation involves a careful choice of constraint equations, here it is just an application of Stokes' theorem changing the contour integral to an area integral
\be
\p^A (\lambda F_{Au})=(\p^A \lambda) F_{Au}+ \lambda \p^A F_{Au}.
\ee
The first term is soft charge while the second term can be evaluated by projecting Maxwell's equations to the boundary
\be
\nabla^\nu F_{\nu\mu}=j_\mu.
\ee
Indeed if we are careful about this we actually get both the massless and massive contributions to the hard charge in one fell swoop! Namely~\cite{Himwich:2019dug}
\be
\gamma^{z\bz}(\p_\bz F_{zu}+\p_z F_{\bz u})=r^2j_u+(\p_u-\p_r)(r^2F_{ru})
\ee
so that the surviving term in the large-$r$ limit involves an area integral of
\be
\int du  J^M_u=\int r^2j_u+(r^2F_{ru})\Big|_{\mathcal{I}^+_\pm}.
\ee
We don't need to add $i^\pm$ in by hand.  To get the full Ward identity for a scattering process we can consider closing the contour to the left or right for the two ways we can cap off the contour on the celestial sphere.  

Our claim is that the charge for the 2D CCFT, defined on a contour $C$, is equal to the canonical charge evaluated on a hypersurface $\Sigma_C$ that lifts this contour into the bulk
\be\label{JCl}
\mathcal{J}_C(\lambda)=\int_{\p\Sigma_C}\star k(\lambda).
\ee
We get the necessary antipodal part (which distinguishes memory effects from a scattering process from plane waves passing through) by nature of the cut through the bulk.  While we don't appear to directly need the antipodal matching to get to the Ward identity, it does come in handy if we want to avoid the bulk.  Namely, as illustrated by the purple surface in figure~\eqref{slice}, we can hug the boundary and jump antipodally across spacelike infinity if we want to consider the flux expressions for the charges. In the case where we think of this surface as hugging the boundary the full Ward identity of in + out is trivially equal to left + right  because we have just split the same boundary $S^3$ differently.

While~\eqref{JCl} is the full charge for this hypersurface, we can isolate a soft contribution by considering a non-constant $\lambda$ and picking a contour that avoids any hard particles. Combining~\eqref{jfuz} and~\eqref{Jcvj}, if we pick a point $w$ where there are no other operator insertions and use Cauchy's integral theorem we see that
\be\label{Jtoj}
\mathcal{J}_{C_w}\left(\lambda=\frac{1}{z-w}\right)=\oint_{C_w}\frac{dz}{2\pi i}\frac{j(z)}{z-w}=j(w).
\ee
Because the soft charge~\eqref{jisqs} only reduced to~\eqref{jfuz} due to the shadow relation, we need to be careful about restricting to a finite region.  However, as we will discuss below, once we introduce the magnetic dual charge we can indeed isolate the single helicity soft contribution in this manner. From the point of view of the bulk, the hypersurface $\Sigma_{C_w}$ corresponding to a small circle around the point $w$ should be a highly boosted image of the $X^3=0$ cut such that it approaches the Rindler horizon for an observer accelerating towards $w$. From the point of view of the extrapolate dictionary we can nominally do this for the massless charge case.  Going to boost eigenstates should help us localize the massive charge contributions towards their respective reference directions.

By contrast only the hard part of the charge survives whenever $\lambda$ is holomorphic inside the contour. As a cute check, the case $\lambda=1$ is indeed an application of Gauss's law for a weird hypersurface
\be\label{kqc}
\int_{\p\Sigma_C}\star k(\lambda)=\sum_{i\in C}Q_i.
\ee
Namely, rather than the constant `ordinary-time' slices we are used to this is a constant Rindler time cut of the bulk (but in the region outside the Rindler wedge). This makes sense when we consider how the hypersurface cuts the worldlines of the particles. The in versus out charges are weighted by a relative sign coming from the orientation of our slice when pushed towards the boundary. From this perspective, the soft factor can be viewed as a Greens function solving~\eqref{kqc}.

\section{Discussion}

We will close with some comments about why this perspective is useful.  First, it is very satisfying to see that the charges in the dual theory correspond to the same symmetries in the bulk. This is closer to how we talk about the relation between bulk and boundary charges in AdS/CFT.  While the choice of hypersurface is unnatural from the point of view of canonical gauge theory, it is natural from the point of view of the celestial dictionary since radial evolution on the sphere corresponds to Rindler evolution in the bulk.  We would like to view the fact that this is clearly not a Cauchy slice as a positive in that it should help inform how to handle issues that might arise in formulating CCFT as a radially quantized theory. 

Second, we see that we are able to work directly with the field strength in our discussion of the large gauge charge.  We only needed the gauge field $A$ to give meaning to the transformation $\lambda$.  If we work in dual variables there is a corresponding magnetic charge and soft theorem~\cite{Strominger:2015bla}.  For the electromagnetic case this is literally
\be
\int_{\p\Sigma} \star k\mapsto \int_{\p\Sigma}  k.
\ee
For a spacelike Cauchy slice the integrand of the surface charge changes from the electric to the magnetic field
\be
r^2 F_{ru}\mapsto  iF_{z\bz}
\ee
for the dual charge.  At null infinity the Hodge dual acts diagonally on the $F_{uz}$ and $F_{u\bz}$ components since these correspond to self dual and anti-self dual solutions, respectively. The curl in~\eqref{QFC} gets replaced by a divergence and we can pick out the single helicity photons by taking the combinations
\be
Q_{(A)SD}=\frac{1}{2}(Q\mp i\widetilde{Q}).
\ee
This alleviates the need to use the shadow relation for the leading soft photon to select out the positive helicity soft theorem.

Finally, recent investigations~\cite{Donnay:2021wrk} have shown that one can split the BMS fluxes into a hard and a soft part which each obey the expected symmetry algebra provided you use the appropriate bracket~\cite{Barnich:2011mi,Campiglia:2020qvc,Compere:2020lrt}.  These involve non-trivial quadratic corrections to the soft generators which one might have anticipated from loop corrections~\cite{He:2017fsb} or from the point of view of the descendancy relations for the celestial diamonds~\cite{Pasterski:2021fjn,Pasterski:2021dqe}.  See also~\cite{Freidel:2021qpz,Freidel:2021dfs}. The picture where we go from the full charge to the soft charge using a reference point~\eqref{Jtoj} away from any hard particles might help explain why we expect such a splitting.  Getting the soft charge in this manner is clearly sensitive to any discussions of boundary terms in the definition of the charge, as well as the (in)ability to isolate hard particles to distinct points on the celestial sphere~\cite{Freidel:2021ytz}.

In sum, we see that taking `radial = Rindler' evolution seriously gives us a nice way to jump directly to the celestial Ward identities from the bulk. The perspective brings us closer the standard holographic relation between between bulk and boundary symmetry generators, informs the sensitivity of these Ward identities to our choice of boundary terms in the charge, and warns us of issues with radial time ordering that we might run into when we consider the fully interacting theory.

\section*{Acknowledgements}
 Many thanks to Laurent Freidel, Sebastian Mizera, and Herman Verlinde for interesting discussions. My research is supported by the Sam B. Treiman Fellowship at the Princeton Center for Theoretical Science.

\bibliographystyle{utphys}
\bibliography{references}

\end{document}